\documentclass[pre,twocolumn,showpacs,preprintnumbers]{revtex4}
\usepackage{graphicx}
\usepackage{dcolumn}
\usepackage{bm}

\begin{document}

\title{van der Waals-like phase separation instability of a driven granular
gas in three dimensions}
\author{Rui Liu, Yinchang Li and Meiying Hou}
\affiliation{Beijing National Laboratory for Condensed Matter Physics, Institute of
Physics, Chinese Academy of Sciences, Beijing 100080, China}
\author{Baruch Meerson}
\affiliation{Racah Institute of Physics, Hebrew University of Jerusalem, Jerusalem 91904,
Israel}
\date{\today }

\begin{abstract}
We show that the van der Waals-like phase separation instability of a driven
granular gas at zero gravity, previously investigated in two-dimensional
settings, persists in three dimensions. We consider a monodisperse granular gas
driven by a thermal wall of a three-dimensional rectangular container at zero
gravity. The basic steady state of this system, as described by granular
hydrodynamic equations, involves a denser and colder layer of granulate located
at the wall opposite to the driving wall. When the inelastic energy loss is
sufficiently high, the driven granular gas exhibits, in some range of average
densities, negative compressibility in the directions parallel to the driving
wall. When the lateral dimensions of the container are sufficiently large, the
negative compressibility causes spontaneous symmetry breaking of the basic
steady state and a phase separation instability. Event-driven molecular dynamics
simulations confirm and complement our theoretical predictions.
\end{abstract}

\maketitle


\section{\label{SecI}Introduction}

Rapid flows of granular materials are widespread in nature and technology.
Losing kinetic energy to microscopic degrees of freedom of the grains in
grain collisions, the granular flows are intrinsically far from thermal
equilibrium and therefore exhibit a host of pattern forming instabilities
\cite{ristow,aranson+tsimring}. Quantitative modeling of granular flows
remains challenging, and pattern forming instabilities can help by providing
sharp tests to the models. In this work we focus on one pattern forming
instability that develops in \textit{granular gas}: an assembly of
inelastically colliding hard spheres. The only dissipative effect in the
particle collisions that we will take into account is a reduction in the
relative normal velocity of colliding particles, accounted for by a
(constant) coefficient of normal restitution $r$. We will assume nearly
elastic collisions, $1-r\ll1$, and small or moderate gas densities. As shown
in many previous studies \cite%
{campbell,kadanoff,thorsten1,thorsten2,goldhirsch1,brilliantov}, these
restrictions make it possible to use equations of granular hydrodynamics.

The phase separation instability, that will be in the focus of our attention
here, was originally predicted from hydrodynamic equations and then observed
in molecular dynamic (MD) simulations in a \textit{two-dimensional} (2D)
setting: a monodisperse gas of inelastically colliding hard disks at zero
gravity, confined in a 2D rectangular box and driven by a side wall that
vibrates with a high frequency and small amplitude \cite%
{livne1,argentina,brey2,khain1,livne2,baruch2,khain2}. The basic steady state of
the 2D system is the \textit{stripe state} \cite{kudrolli,grossman}: a stripe of
a denser and colder gas located at the wall opposite to the driving wall. At
sufficiently high energy loss, and within a certain (``spinodal") interval of
grain area fractions, the stripe state becomes unstable with respect to small
density perturbations in the lateral
direction, unless the lateral container size is too small \cite%
{livne1,brey2,khain1}. Within a broader binodal, or coexistence, interval,
the stripe state is \textit{metastable} \cite%
{argentina,khain2}. In both cases one finally observes, usually after a
coarsening process, a granular ``drop" coexisting with ``vapor", or a granular
``bubble" coexisting with ``liquid", along the wall opposite to the driving wall
\cite{argentina,livne2,khain2}. This remarkable far-from-equilibrium
two-dimensional (2D) phase separation phenomenon is in many ways similar to the
gas-liquid transition as described by the classical van der Waals equation of
state, but the role of temperature is now played by the inelastic energy loss,
see below.   The basic properties of the phase separation in 2D are
qualitatively captured by an effective one-dimensional granular hydrodynamic
model, suggested in Ref. \cite{cartes}. Recently, the studies of the phase
separation in 2D have been extended to an annular geometry \cite{manuel}.

The present work predicts a similar phase separation phenomenon \textit{in three
dimensions} (3D). By extending the previous treatments to 3D we are breaking
ground for a future investigation of this phase separation phenomenon in reduced
gravity experiments. The paper is organized as follows. In Sec. \ref{SecII} we
employ a hydrodynamic description to describe the ``layer state" (the basic
state of the system), to compute the spinodal balloon and the binodal asymptote
of the system, and to determine the critical lateral dimensions of the container
for the phase separation to occur. As this hydrodynamic description will be
dealing only with steady states with a zero mean flow, the corresponding theory
may be called hydrostatic. In Section \ref{SecIII} we report a series of
event-driven MD simulations of this driven granular system and compare the
simulation results with the hydrostatic theory predictions. Section \ref{SecIV}
summarizes our results, discusses possible morphologies of phase-separated
states and briefly mentions some extensions of the model.

\section{\label{SecII}Granular hydrostatics: the layer state and the
phase separation}

\subsection{The density equation}

Let $N$ hard spheres of diameter $d$ and mass
$m=1$ move, at zero gravity, inside a rectangular container with dimensions $%
L_x$, $L_y$ and $L_z$. The spheres collide inelastically with a constant
coefficient of normal restitution $r$. For simplicity, we neglect the
rotational degree of freedom of the particles. Let one of the container
walls perform high-frequency and small-amplitude vibrations in the $x$%
-direction. We assume that the vibration amplitude is much less than the mean
free path of the particles at the driving wall, while the vibration frequency is
sufficiently high. In this case one can treat the vibrating wall as effectively
immobile and prescribe a constant gas temperature $T_0$ at this wall
\cite{knudsen}. For simplicity, particle collisions with all other walls of the
container are considered elastic. The energy transferred from the ``thermal"
wall to the granulate dissipates in the inter-particle collisions, and we assume
that the system reaches a time-independent state with a zero mean flow. In the
nearly elastic limit, $1-r \ll 1$, and for small or moderate particle densities,
one can safely use granular hydrodynamic equations. For a zero-mean-flow steady
state these are reducible to two \textit{hydrostatic} relations:
\begin{equation}
\nabla \cdot \left[\kappa \nabla T(\mathbf{r}) \right] =I\,, \; \; \;
p=const\,,  \label{en_balance}
\end{equation}
where the gas pressure $p=p(n,T)$, heat conductivity $\kappa=\kappa (n,T)$
and energy loss rate due to inelastic collisions $I=I(n,T)$ depend on the
particle number density $n(\mathbf{r})$ and granular temperature $T(\mathbf{r%
})$ of the gas. We will employ the equation of state of Carnahan and
Starling \cite{carnahan} and the widely used semi-empiric transport
coefficients derived from kinetic theory in the spirit of Enskog approach
\cite{jenkins}:
\begin{eqnarray}
p &=& n T(1+4G_0) \,,  \label{pressure} \\
\kappa &=& \frac{4dnT^{1/2}G_0}{\sqrt{\pi}} \left[1+\frac{9\pi}{32}\left(1+%
\frac{5}{12 G_0}\right)^{2} \right]\,,  \label{heatcond} \\
I &=& \frac{12(1-r^2)nT^{3/2}G_0}{\sqrt{\pi}d}\,,  \label{loss}
\end{eqnarray}
where
\[
G_0(\nu)=\frac{\nu (1-\nu /2)}{(1-\nu)^{3}}\,,
\]
and $\nu= \pi d^{3} n/6$ is the local value of the solid fraction of the
grains. Let us rescale all the coordinates by $L_x$ and introduce the
rescaled inverse density $u(\mathbf{r})=n_{c}/n(\mathbf{r})$, where $n_{c}=%
\sqrt{2}/d^3$ is the crystal packing density in 3D. The rescaled coordinates
$x,\,y$ and $z$ now run between zero and $1,\,\Delta_y=L_y/L_x$, and $%
\Delta_z=L_z/L_x$, respectively, while Eqs.~(\ref{en_balance})-(\ref{loss})
can be transformed, after some algebra, into a single equation for the
inverse density $u(\mathbf{r})$:
\begin{equation}
\nabla \cdot \left[ F(u)\nabla u\right] =\Lambda Q(u)\,,
\label{eq:density_eq}
\end{equation}%
where $F(u)=A(u) B(u)$,
\begin{eqnarray}
A(u)&=&\frac{G}{u^{1/2}(1+4G)^{5/2}} \left[ 1+ \frac{9\pi}{32} \left(1+\frac{%
5}{12G}\right)^{2} \right] \,,  \nonumber \\
B(u)&=&1+4G+\frac{4\pi}{3\sqrt{2}} \frac{u \left[u \left(u+\frac{\pi}{3\sqrt{%
2}}\right) - \left(\frac{\pi}{6}\right)^{2}\right]} {\left(u-\frac{\pi}{3%
\sqrt{2}}\right)^{4}} \,,  \nonumber \\
Q(u)&=&\frac{9}{\pi} \frac{u^{1/2}G}{(1+4G)^{3/2}}\,,  \nonumber \\
G(u)&=&\frac{\pi}{3\sqrt{2}}\frac{u\left(u-\frac{\pi}{6\sqrt{2}}\right)} {%
\left(u-\frac{\pi}{3\sqrt{2}}\right)^{3}}\,,
\end{eqnarray}
while $\Lambda =(\pi /3)(1-r^2)(L_x/d)^2$ is the dimensionless hydrodynamic
inelastic loss parameter. The boundary conditions for Eq.~(\ref%
{eq:density_eq}) are the zero heat flux conditions $\partial_n u =0$ at all
walls except the thermal wall $x=1$, and the condition $\partial_y u =
\partial_z u=0$ at $x=1$. The latter condition follows from the constancy of
the temperature at the thermal wall \cite{knudsen}, combined with the constancy
of the pressure in a steady state. As the total number of particles $N$ is
conserved, we obtain
\begin{equation}
\frac{1}{{\Delta}_y {\Delta}_z}\int_{0}^{1}dx\int_{0}^{{\Delta}%
_y}dy\int_{0}^{{\Delta}_z} \frac{dz}{u(x,y,z)}=f\,,  \label{eq:norma}
\end{equation}
where $f=\langle n \rangle/{n}_c$ is the average volume fraction of the
granulate, and $\langle n \rangle=N/(L_x L_y L_z)$ is the average number density
of the particles in the container. The nonlinear partial differential equation
(PDE) (\ref{eq:density_eq}), together with the boundary conditions and the
normalization condition (\ref{eq:norma}), determine all possible steady state
density profiles, governed by two dimensionless parameters $f$ and $\Lambda$.
The density profiles are independent of $T_0$.
Importantly, at large $u$ the function $Q(u)$ decreases with an increase of $%
u$. This implies that the steady state solution of Eq.~(\ref{eq:density_eq})
is non-unique \cite{non-unique} which paves the way to phase separation and
coexistence, as in the 2D case.

\subsection{The layer state, spinodal balloon and binodal asymptote}

\begin{figure}[htp]
\includegraphics[width=6.5cm]{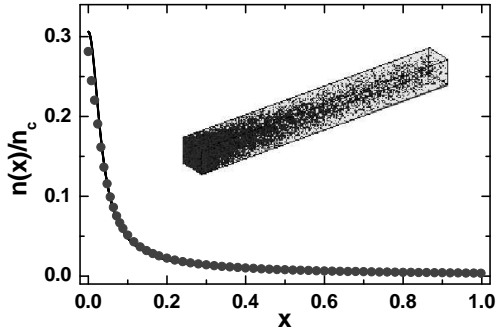}
\caption{The rescaled density profiles obtained from the hydrostatic
equations (line) and from MD simulations (circles). The dimensionless
parameters are $\Lambda=5 \times 10^3$ and $f=0.02317$, the MD simulation
parameters are $L_x = 500d$, $L_y = L_z = 50d$, $N=40960$ and $r= 0.9904$.
The inset shows a typical snapshot of the system at the steady state as
observed in the MD simulation.}
\label{fig1}
\end{figure}

The simplest solution of Eq.~(%
\ref{eq:density_eq}) $u= U(x)$ is laterally symmetric, that is independent
of $y$ and $z$. This is the ``layer state", and it is fully determined by
the following equations:
\begin{equation}
[F(U) U^{\prime}]^{\prime} = \Lambda Q(U),\; U^{\prime}(0)=0, \;
\int_{0}^{1} \frac{dx}{U(x)}=f,  \label{eq:stripe}
\end{equation}
where the primes denote the $x$-derivatives. Figure~\ref{fig1} depicts an
example of the rescaled density profile of the layer state obtained by
solving Eqs. (\ref{eq:stripe}) numerically. The hydrostatic density profile
agrees with a late-time profile observed in our MD simulations described
below.
\begin{figure}[htp]
\includegraphics[width=6.5cm]{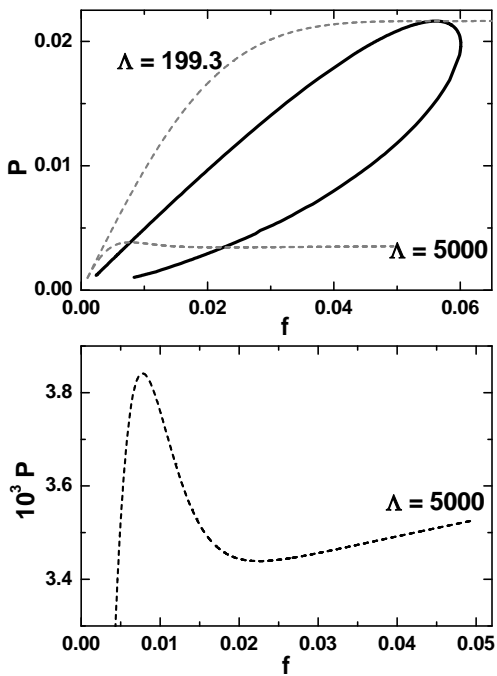}
\caption{Upper panel: the rescaled steady state granular pressure $P$ versus the
grain volume fraction $f$ for $\Lambda=199.3$ and $\Lambda=5\times 10^{3}$.
Shown by the solid line is the spinodal balloon, inside which the effective
lateral compressibility of the gas is negative. The borders $f_1$ and $f_2$ of
the spinodal interval are determined from the condition $\left( \partial
P/\partial f \right)_{\Lambda}=0$. Lower panel: a zoom-in at the $P(f)$
dependence for $\Lambda=5\times 10^{3}$.} \label{fig2}
\end{figure}

\begin{figure}[htp]
\includegraphics[width=6.5cm]{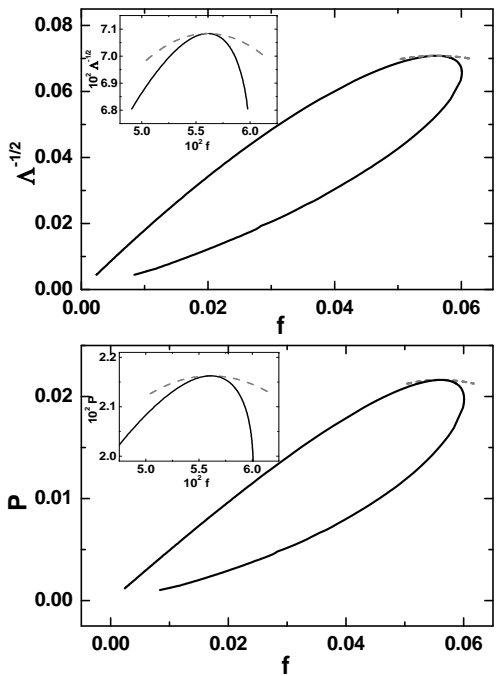}
\caption{The spinodal balloon of the system (the solid line) and the asymptote
of the binodal line in a close vicinity of the critical point (the dashed line)
on the plane $f,\Lambda^{-1/2}$ (the upper panel) and on the plane $f,P$ (the
lower panel). The insets zoom in on a close vicinity of the critical point.}
\label{fig3}
\end{figure}

Having found the density profiles at different $\Lambda$ and $f$, we can
compute, with the help of Eq.~(\ref{pressure}), the rescaled pressure of the
layer state $P=p/(n_{c}T_{0})$. As the steady-state pressure is constant
throughout the system, one can compute it at the thermal wall $x=1$, where the
temperature $T=T_{0}$ is prescribed \cite{knudsen}. We obtain, therefore,
\[
P(f,\Lambda)=\frac{1+4 G[U(1)]}{U(1)}\,.
\]
Two typical $P(f)$ curves for different $\Lambda$ are shown in Fig.~\ref%
{fig2}. At small $\Lambda$ (exemplified by $\Lambda=199.3$) the bulk energy loss
is not very important, and $P(f)$ is monotone increasing with $f$. At
sufficiently large $\Lambda$ (exemplified by $\Lambda=5\times 10^3$) there is an
interval of the volume fractions $f_{1}(\Lambda)<f<f_{2}(\Lambda)$ where the
rescaled pressure $P(f)$ \textit{decreases} with an increase of $f$. Therefore,
the effective compressibility of the gas in the lateral directions is negative
there. The lower panel of Fig.~\ref{fig2} shows a blowup of the negative
compressibility region at $\Lambda=5\times 10^3$. The borders of the negative
compressibility region are determined by the condition $\left(\partial
P/\partial f\right)_{\Lambda}=0$. By joining the spinodal points $f_1$ and
(separately) $f_2$ at different $\Lambda$, we
obtain the spinodal balloon of the system in the $(P,f)$ plane (Fig.~\ref%
{fig2}), or in the $(\Lambda,f)$ plane (Fig.~\ref{fig3}). As $\Lambda$
decreases, the spinodal interval $f_{1}(\Lambda)<f<f_{2}(\Lambda)$ shrinks
into a point, as in the 2D case \cite{argentina,khain2}. This is the
critical point of the system $(P_{c},f_c)$, or $(\Lambda_{c},f_c)$. At $%
\Lambda<\Lambda_c$ $P(f)$ is monotone increasing.

A negative lateral compressibility implies that, within the spinodal
balloon, the layer state [a 1D solution of the steady state equation (\ref%
{eq:density_eq})] is unstable with respect to small-amplitude long-wavelength
perturbations in one or both lateral directions. Similarly to the well-studied
2D setting \cite{argentina,khain2} there is also a binodal (or coexistence)
line, originating from the intervals of area fractions where the layer state,
although linearly stable, is nonlinearly unstable (that is, metastable). The two
branches of the binodal line in the $(\Lambda,f)$ plane merge at the same
critical point $(\Lambda_{c},f_c)$. The asymptote of the binodal line in a close
vicinity of the critical point, $|f-f_c| \ll f_c$ and $0<\Lambda-\Lambda_c \ll
\Lambda_c$, can be readily established, \textit{cf.} Refs.
\cite{argentina,khain2}. Indeed, in the close vicinity of the critical point
$P(f)$ is describable, at fixed $\Lambda$, by a cubic parabola in $f-f_c$
without a quadratic term. As a result, one can find, at fixed $\Lambda$, the two
points $f_{-}$ and $f_{+}$, belonging to the binodal line, from the simple
relations $P(f_{-},\Lambda)=P(f_{+},\Lambda)$ and $f_{-}+f_{+}=2 f_c$. The
resulting binodal asymptote is depicted in Fig.~3.

Unfortunately, this simple asymptote cannot be continued beyond the close
vicinity of the critical point. The form of the binodal line far from the
critical point has not yet been derived from granular hydrodynamics, neither in
2D, nor in 3D. Such a derivation would require a non-perturbative solution of
the nonlinear PDE~(\ref{eq:density_eq}). Most likely, this can only be done
numerically. Note that the ``Maxwell construction", suggested in
Ref.~\cite{argentina} for the binodal line in 2D, is valid only in a close
vicinity of the critical point, where it is reducible to the two simple
relations $P(f_{-},\Lambda)=P(f_{+},\Lambda)$ and $f_{-}+f_{+}=2 f_c$. The
reader is advised to consult with Ref.~\cite{khain2} for a more detailed
discussion of this issue.

\subsection{The critical value of lateral dimensions: a marginal stability analysis}

When the dimensionless parameters $\Lambda$ and $f$ are within either spinodal,
or binodal balloon, a steady state with a broken lateral symmetry should
develop. However, the phase separation demands a sufficiently large lateral size
of the system. It will be suppressed by the lateral heat conduction if the
lateral aspect ratios $\Delta_y$ and $\Delta_z$ are both less than a critical
value $\Delta_c(\Lambda,f)$. By analogy with 2D, see
Refs.~\cite{livne1,khain1,livne2,baruch2}, we can determine
$\Delta_c(\Lambda,f)$ from a marginal stability analysis. Indeed, let $\Delta_y$
and $\Delta_z$ be less than $\Delta_c(\Lambda,f)$, so the layer state is
linearly stable, because of the lateral heat conduction, even within the
spinodal balloon. Increasing $\Delta_y$ and/or $\Delta_z$ slightly beyond
$\Delta_c(\Lambda,f)$, one should observe a (weakly) phase separated steady
state that bifurcates supercritically from the layer state. Therefore, close to
the bifurcation point, this weakly phase separated steady state can be found by
linearizing Eq.~(\ref{eq:density_eq}) around the layer state  $u=U(x)$. In the
time-dependent hydrodynamic framework, this linear analysis corresponds to a
\textit{marginal stability} analysis of the layer state with respect to small
perturbations in the $y$- and $z$-directions.

\begin{figure}[htp]
\includegraphics[width=6.5cm]{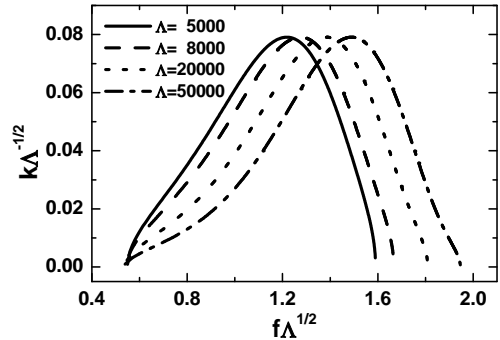}
\caption{Marginal stability curves for four different values of $\Lambda$ are
plotted in rescaled coordinates: $k\Lambda^{-1/2}$ versus $f\Lambda^{1/2}$.  For
a fixed $\Lambda$ the layer state is linearly stable above the corresponding
curve and unstable below the curve.} \label{marginal}
\end{figure}

Substituting $u(x,y,z) = U(x)+\psi_k(x)\cos k_y y \,\cos k_zz$ into
Eq.~(\ref{eq:density_eq}) and linearizing  with respect to the small correction
$\psi_k(x)\cos k_y y \cos k_z z$, we obtain the following linear equation for
the new function $\phi_k(x)\equiv F \,\psi_k(x)$:
\begin{equation}
\phi_k^{\prime\prime}(x)-\left(k^2+\frac{\Lambda Q_U}{F}\right)\phi_k (x)= 0\,.
\label{E8}
\end{equation}
Here $k^2=k_y^2+k_z^2$, the functions $F$ and $Q$ are evaluated at $u=U(x)$, and
$Q_U$ denotes the $u$ derivative of $Q(u)$ evaluated at $u = U(x)$. The boundary
conditions for Eq.~(\ref{E8}) are
\begin{equation}
\phi_k^{\prime}(x=0)=0\,\;\;\;\;\mbox{and}\;\;\;\;\;\;   \phi_k(x=1)=0\,.
\label{E9}
\end{equation}
Equation~(\ref{E8}) can be interpreted as a Schr\"{o}dinger equation for a
single particle in a one-dimensional potential $V(x)=\Lambda Q_U/F$, while the
quantity $-k^2$ serves as the (negative or zero) energy eigenvalue. We solved
this eigenvalue problem numerically for different $\Lambda$ and $f$.
Figure~\ref{marginal} shows the resulting marginal stability curves
$k_{*}=k_{*}(f)$ for four different values of $\Lambda>\Lambda_c$. Assuming that
the instability is non-oscillatory at the onset, one can interpret the marginal
stability results as linear instability of the layer state  below the
corresponding curve, and linear stability above the curve. The instability is
possible only within the spinodal balloon: the borders $f_1$ and $f_2$ of the
spinodal interval correspond, at fixed $\Lambda$, to zero eigenvalues $k\to 0$.

It can be seen in Fig.~\ref{marginal} that, in the rescaled coordinates, the
marginal stability curves for different $\Lambda$ originate (almost) at the same
point of the horizontal axis $f \Lambda^{1/2}$. Furthermore, the maxima of all
the curves are almost equal. Like in the 2D case, the first property can be
explained analytically by considering the dilute limit of the problem, while the
second property results from the strong localization of the eigenfunctions
$\phi_k(x)$ near the elastic wall \cite{khain1}.

Having found the eigenvalues $k_{*}(f,\Lambda)$, we can determine the critical
(minimum) lateral aspect ratios for a phase separation. Indeed, the zero heat
flux conditions at the walls $y=0,\,y=\Delta_y,\,z=0$ and $z=\Delta_z$ (recall
that we are using rescaled coordinates) yield the quantization rules
$k_y=(n_y\pi)/\Delta_y$ and $k_z=(n_z\pi)/\Delta_z$, where $n_y,n_z=0,1,2,
\dots$. Therefore, the minimum value of $\Delta_y$ ($\Delta_z$) for a phase
separation only in the $y$-direction (correspondingly, only in the
$z$-direction) is $\Delta_y^{c}=\Delta_z^{c}=\pi/k_{*} (f, \Lambda)$. For
example, for $\Lambda=8 \times 10^3$ and $f=0.011$ we obtain $k^2_* \simeq
21.0$, therefore $\Delta_y^{c}=\Delta_z^{c}=\pi/k_{*} \simeq 0.69$. In order to
have a phase separation in \textit{both} lateral directions $y$ and $z$, the
aspect ratios $\Delta_y$ and $\Delta_z$ must obey the inequality
$$
\frac{1}{\Delta_y^2}+\frac{1}{\Delta_z^2}< \frac{k_{*}^2 (f, \Lambda)}{\pi^2}\,.
$$

\section{\label{SecIII}MD Simulations}

\subsection{Method}

We performed a series of event-driven MD simulations of
this 3D system using a standard algorithm described by Rapaport \cite%
{rapaport}. Simulations involved $N$ hard spheres of diameter $d=1$ and mass
$m=1$. After each collision of particle $i$ with particle $j$, their
relative velocity was updated according to
\begin{equation}
\vec{v}_{ij}^{\,\prime}=\vec{v}_{ij} - \left( 1+r \right) \left( \vec{v}%
_{ij}\cdot \hat{r}_{ij}\right)\hat{r}_{ij}\,,  \label{eq:velocs}
\end{equation}
where $\vec{v}_{ij}$ is the precollisional relative velocity, and $\hat{r}%
_{ij}\equiv \vec{r}_{ij}/\left|\vec{r}_{ij}\right|$ is a unit vector
connecting the centers of the two particles. The ``thermal" wall was kept at
constant temperature $T_{0}$ that we set to unity. We used a standard
thermal wall implementation, see \textit{e.g.} Ref. \cite{thorsten3}, p.
173-177. Particle collisions with the rest of the walls were assumed
elastic. The natural time unit of the MD simulations is $d(m/T_{0})^{1/2}=1$%
. The initial spatial distribution of (non-overlapping) particles was
uniform in all simulations. The initial particle velocity distribution was
uniform in the direction angles, while the absolute value $v_0$ of the
velocity of each particle was chosen to be such that $(1/2)\,m v_0^2=(3/2)
(1-1/N) T_0$. In all simulations the velocity of the center of mass of the
particles at $t=0$ was zero. That the transients died out and the system
reached a steady state was monitored by (i) measuring the total energy of
all particles versus time, and (ii) measuring the coordinates of the center
of mass versus time.

As a test simulation, we performed a simulation of a system which
dimensionless parameters $\Lambda$ and $f$ are within the spinodal balloon
but which still cannot phase separate because of too small lateral
dimensions, see Fig.~\ref{fig1}. The inset shows a late-time snapshot of the
system as observed in the MD simulation. As one can see from Fig.~\ref{fig1}%
, the measured particle number density, rescaled to $n_c$, as a function of
the rescaled distance from the driving wall is in good agreement with our
hydrostatic calculations.

\subsection{Phase separation}

\begin{figure}[tph]
\includegraphics[width=6.5cm]{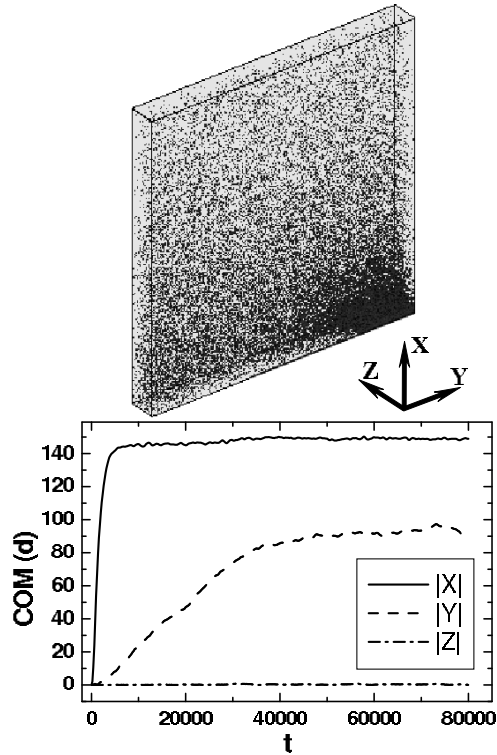}
\caption{Upper panel: a late-time snapshot of an MD simulation with $%
N=10^{5} $ particles of $r=0.89945$ in the container with dimensions $%
L_{x}=L_{y}=500d $ and $L_{z}=50d$. The upper wall is the driving wall. The
dimensionless parameters of the system $\Lambda =5\times 10^{4}$ and $%
f=0.0057$ correspond to a point within the spinodal balloon of Fig.~\protect
\ref{fig2}. Lower panel: the absolute values of the three center-of-mass (COM)
coordinates (measured in the units of the particle diameter $d$) versus time
[measured in the units of $d(m/T_{0})^{1/2}$]. The center of the container is at
the origin.} \label{fig4}
\end{figure}

Remaining within the spinodal balloon, and increasing \textit{one} of the
lateral dimensions of the system, we observed phase separation as expected from
the theory, see Fig.~\ref{fig4}. A dense cluster develops in one of the two
corners of this quasi-2D Hele-Shaw cell, at the wall opposite to the driving
wall. Quantitative diagnostics are provided by the plots of the three
center-of-mass coordinates of the system versus time, shown in the lower panel
of Fig.~\ref{fig4}.

\begin{figure}[tph]
\includegraphics[width=6.5cm]{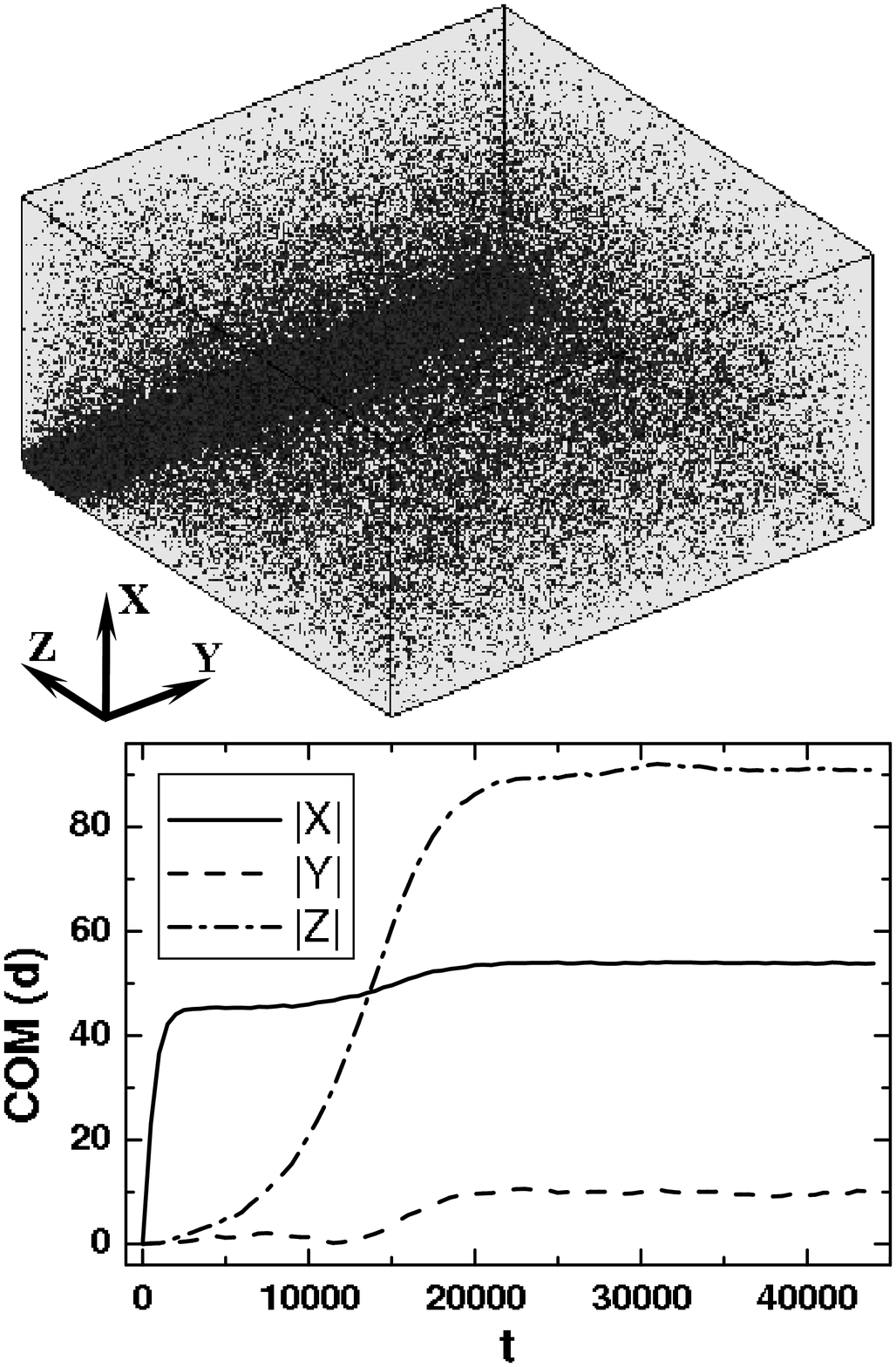}
\caption{Upper panel: a late-time snapshot of an MD simulation with $%
N=5\times 10^{5}$ particles of $r=0.899452$ in the container with dimensions
$L_{x}=200d$ and $L_{y}=L_{z}=400d$. The upper wall is the driving wall. The
dimensionless parameters of the system $\Lambda =8\times 10^{3}$ and $%
f=0.011 $ correspond to a point within the spinodal balloon of Fig.~\protect
\ref{fig2}. Lower panel: same as in Fig.~\protect\ref{fig4}.}
\label{fig5}
\end{figure}

\begin{figure}[htp]
\includegraphics[width=6.5cm]{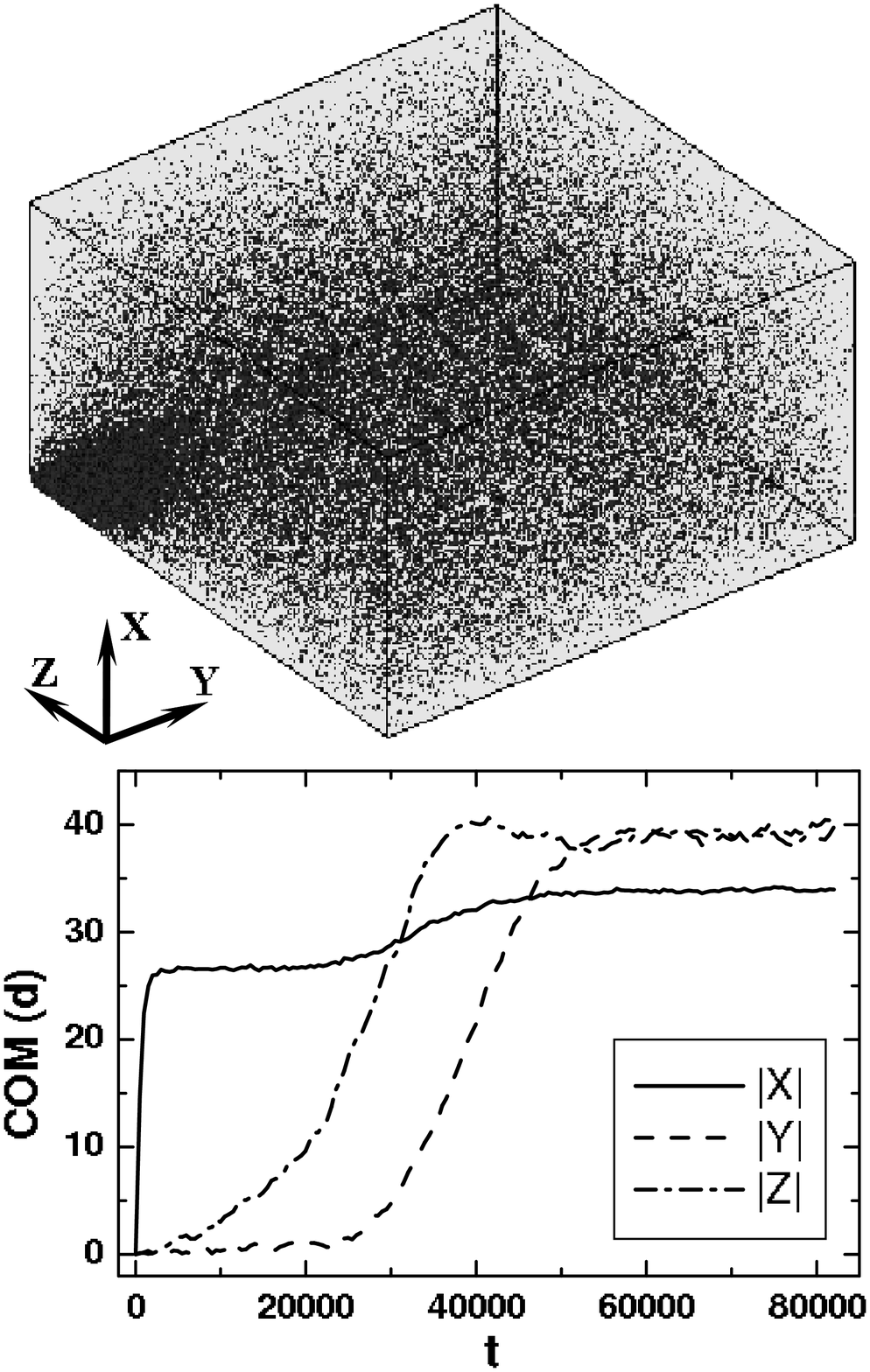}
\caption{Upper panel: a late-time snapshot of an MD simulation with $%
N=340736 $ particles of $r=0.899452$ in the container with dimensions $%
L_x=200d$ and $L_y = L_z=400d$. The upper wall is the driving wall. The
dimensionless parameters of the system $\Lambda = 8\times 10^3$ and $f=7.529
\times 10^{-3} $ correspond to a point within the spinodal balloon of Fig.~%
\protect\ref{fig2}. Lower panel: same as in Fig.~\protect\ref{fig4}.}
\label{fig6}
\end{figure}

Figure~\ref{fig5} shows another example of phase separation for $\Lambda$
and $f$ within the spinodal balloon, but this time in the case when \textit{%
both} lateral dimensions $L_y$ and $L_z$ are sufficiently large. As one can
see, a dense stripe-like cluster forms along one of the edges of the wall
opposite to the driving wall.

Both Fig.~\ref{fig4}, and Fig.~\ref{fig5} show phase separated states with a 2D,
rather than 3D, structure. An example of a fully 3D structure is shown in
Fig.~\ref{fig6}. Here a fully 3D dense cluster (a ``drop") develops in one of
the corners at the wall opposite to the driving wall.

\section{\label{SecIV}Discussion}

As we have shown, granular hydrodynamics predicts negative lateral
compressibility and, therefore, phase-separation instability of the basic state
of a granulate driven by a thermal wall of a rectangular container at zero
gravity. When the lateral dimensions of the container are sufficiently large,
the negative compressibility causes a van der Waals-like phase separation
instability.

Our MD simulations are in agreement with hydrostatic predictions (of course, if
we disregard small fluctuations caused by the discreteness of the particles). In
the language of hydrostatics, a broken-symmetry steady state is described by
either a 2D, or a fully 3D solution of the nonlinear partial differential
equation (\ref{eq:density_eq}) subject to the fixed mass constraint
(\ref{eq:norma}) and the boundary conditions. Such steady-state solutions can be
obtained only numerically (see Ref. \cite{livne1} for 2D examples). Because of
the translational symmetry of the steady-state equations in the $y$ and $z$
directions, bounded solutions satisfying the no-flux boundary conditions in
these directions, must be either independent of the $y$ and $z$ coordinates, or
periodic in them. Furthermore, by analogy with the 2D setting, one should expect
that dynamic coarsening selects a periodic steady state solution with a
\textit{maximum} spatial period (equal
to \textit{twice} the container size in the corresponding direction \cite%
{livne2}).

When one of the lateral aspect ratios, say $\Delta_y$, is larger than the
critical value $\Delta_c(\Lambda,f)$, while the other one, $\Delta_z$, is
smaller than $\Delta_c$, a 2D pattern should develop, and this is indeed
what we observed in Fig.~\ref{fig4}. When both of the lateral dimensions are
larger than $\Delta_c$, we observed dense clusters of two different
morphologies: either a 2D morphology, like the one shown in Fig.~\ref{fig5},
or a fully 3D morphology, like the one shown in Fig.~\ref{fig6}.

What will happen when the parameters $\Lambda$ and $f$ are within the
spinodal or binodal balloons, and its lateral dimensions $L_x$ and $L_y$ are
\textit{much} larger than $\Delta_C$? We expect that \textit{multiple}
``drops" (or bubbles) will nucleate at the wall opposite to the driving wall
and undergo dynamic coarsening, qualitatively similar to Ostwald ripening
\cite{Ostwald}, before reaching the final state with a single drop (or
bubble). The Ostwald ripening regime is beyond the reach of our present
computing resources.

When the lateral aspect ratios $\Delta_y$ and $\Delta_z$ are just above (or
just below) the critical value $\Delta_c=\Delta_c(\Lambda,f)$, the phase
separation, as predicted by hydrostatics, should be ``weak" and look, in the
supercritical case, as a small-amplitude modulation of the layer state \cite%
{livne1,brey2,khain1,livne2,baruch2,khain2}. In 2D such a system experiences
large fluctuations \cite{baruch2}, and it would be interesting to find out
whether large fluctuations persist in 3D.

We also performed a series of MD simulations for more realistic conditions,
using a (truly) vibrating wall instead of the thermal wall, and allowing for
inelastic particle collisions with the walls. Qualitatively, the results have
not changed: for sufficiently high inelasticity of particle-particle collisions
we observed phase separation for intermediate values of the volume fraction, and
no phase separation for too a small or too a large volume fraction.

Our theory and simulations assumed a a zero gravity. To what extent is the van
der Waals-like phase transition sensitive to the presence of a small gravity,
with acceleration $g$, directed towards the thermal wall? In a 2D setting this
question was addressed by Khain and Meerson \cite{khainconv}. Assuming dilute
limit of granular hydrodynamics, they found that, as the Froude number
$F=mgL_x/T_0$ increases, the phase separation crosses over to ``thermal"
granular convection. One can expect a similar scenario in 3D as well, though
this question has not been yet been considered in detail.

In summary, the van der Waals-like phase separation in 2D and 3D provides a
useful and rich prototypical model system for testing the ideas and methods of
granular dynamics. By focusing our attention on 3D in this work, we broke ground
for a future investigation of this fascinating phase separation phenomenon in
reduced gravity experiments.

\begin{acknowledgments}
MH acknowledges financial support from the Chinese National Science
Foundation (grant No. 0402-10474124) and from Chinese Academy of Sciences
(grant No. KACX2-SW-02-06). BM acknowledges financial support from the
Israel Science Foundation (grant No. 107/05) and from the German-Israel
Foundation for Scientific Research and Development (Grant I-795-166.10/2003).
\end{acknowledgments}


\begin{thebibliography}{99}
\bibitem{knudsen} In the Knudsen layer at the thermal wall
the gas temperature is actually smaller than the wall temperature. This is the
well-known Knudsen effect that also occurs in ordinary gases \cite{chapman}. The
Knudsen effect is small once the size of the Knudsen layer (which is of the
order of the mean free path of the gas near the thermal wall) is much less than
the characteristic length scale of the hydrostatic profile. Our hydrostatic
model neglects the Knudsen effect.
\end{thebibliography}
\end{document}